\pdfoutput=1
\documentclass{article}

\usepackage[utf8]{inputenc} 
\usepackage[T1]{fontenc}    
\usepackage{hyperref}       
\usepackage{url}            
   
\usepackage{amsfonts}
\usepackage{amsmath,amssymb}
\usepackage{mathtools}      

\usepackage{graphicx}
\usepackage{tabularx}
\usepackage{subcaption}

\usepackage[ruled, vlined, linesnumbered]{algorithm2e}

\title{Group based Centrality for Immunization of Complex Networks }

\author{
  Chandni ~Saxena \\
  \texttt{cmooncs@gmail.com} 
   \and
M.N.~Doja \\
  \texttt{mdoja@jmi.ac.in} 
   \and
Tanvir ~Ahmad\\
\texttt{tahmad2@jmi.ac.in } \\ 
\and 
 Department of Computer Engineering,
 \\Jamia Millia Islamia, 
 \\New Delhi-25, India }

\begin{document}
\maketitle

\begin{abstract}
Network immunization is an extensively recognized issue in several domains like virtual network security, public health and social media, to deal with the problem of node inoculation so as to minimize the transmission through the links existed in these networks. We aim to identify top ranked nodes to immunize networks, leading to control the outbreak of epidemics or misinformation. We consider group based centrality and define a heuristic objective criteria to establish the target of key nodes finding in network which if immunized result in essential network vulnerability. We propose a group based game theoretic payoff division approach, by employing Shapley value to assign the surplus acquired by participating nodes in different groups through the positional power and functional influence over other nodes. We tag these key nodes as Shapley Value based Information Delimiters (SVID). Experiments on empirical data sets and model networks establish the efficacy of our proposed approach and acknowledge performance of node inoculation to delimit contagion outbreak. 
\end{abstract}


\section{Introduction}
The last decade acknowledged tremendous growth of the array of networks in different real world complex systems and also flourished the debatable issues related to these networks to understand them. Numerous works have attempted to reform geometrical statistics, performance, dynamical behavior and robustness of such networks. A crucial dynamical mechanism in these complex networks is the spreading. This spreading phenomenon pertinent to our everyday life shows up such dynamics as: epidemic on social networks, virus propagation on technological networks or cascading extinctions in food web\cite{Bishop_2011,Salath__2010}. These completely different dynamics can be modeled as the propagation on set of nodes and links through interactions. A set of structural nodes, which if initiated can spread the information to the entire network, else if immunized can block the effective diffusion to the largest of this network. There has been substantial interest to determine influential spreaders in the network, to be the target of immunization efforts in order to assist the network in contact with epidemics\cite{Singh_2015,Shams_2013}. A well-established approach, also known as targeted immunization to protect the network against such threat of spread is to identify a set of critical nodes, if removed/immunized would result in essential network vulnerability\cite{Singh_2014}.

\begin{figure}[!h]
\begin{center}
\includegraphics[width=0.6\linewidth]{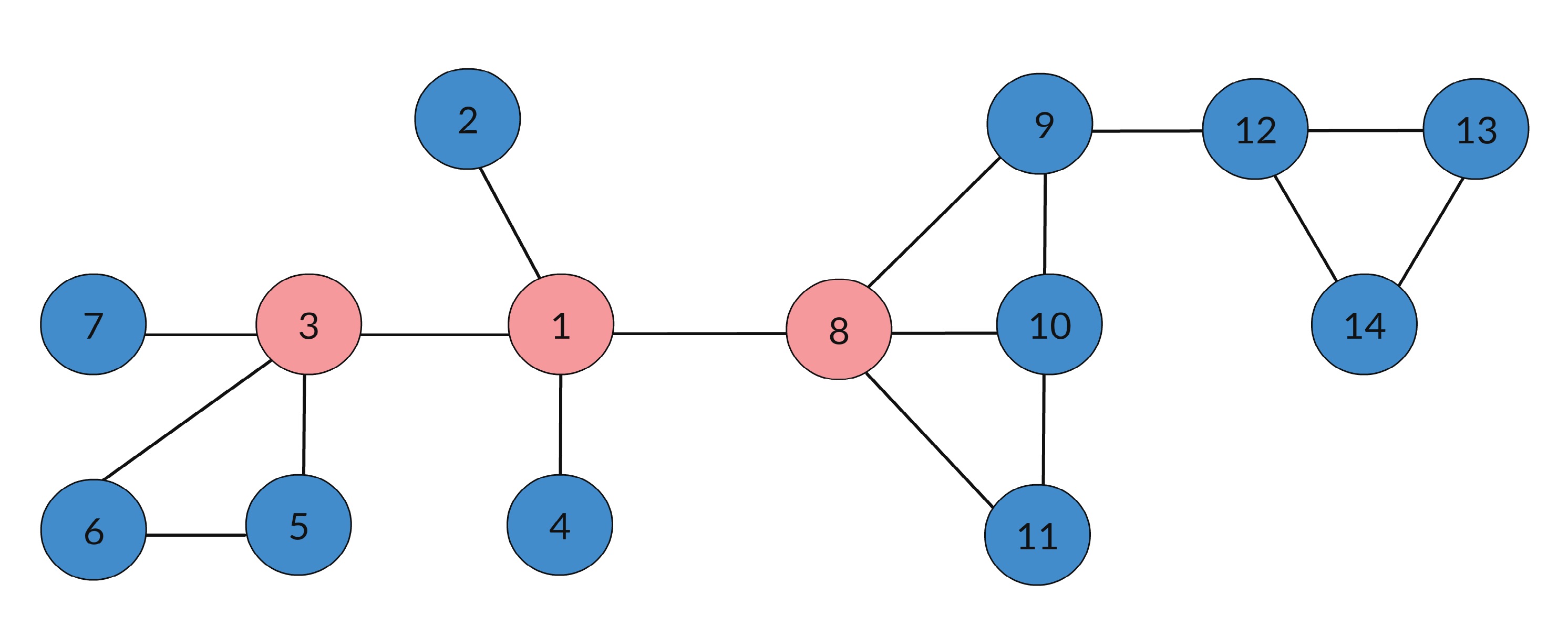}
\caption{A schematic network}
\label{fig. 1}
\end{center}
\end{figure}

The concept of centrality heuristic built upon topological properties of the network can be significant to quantify the importance of critical nodes. Clearly, the conventional centralities\cite{Zweig_2016} bring out different characteristics of the nodes depending upon the objective criteria of these measures. Degree centrality identifies highly connected nodes also called hubs with greatest neighboring sphere of influence. Immunizing highest degree nodes leads to effective reduction in the count of infected nodes in network spreading. The highest degree nodes immunization results in adequate reduction in network density which affect epidemic growth rate. The closeness centrality chooses the best nodes in the terms of fastest information flow with least average propagation length in the network. Hence, targeting immunization on nodes with high value of this measure may lead to increase in the average paths length. Further, betweenness centrality describes nodes in terms of potential power in controlling information flow with maximum average number of shortest paths passing through it. Therefore immunization of nodes based on this metric eliminates the number of transmission routes for spreading. All these strategies are efficient targeted immunization as compared to random immunization which is based on selecting a random subset of k nodes\cite{Cohen_2001}. However, it does not provide the optimal solution for immunization problem because a greedy algorithm selects the k optimal nodes individually for the target in place of selecting optimal set of k nodes\cite{Borgatti}.\\ Standard centrality measures access the goodness of a node by unifying the important role played by the node itself. Considering the schematic network in \ref{fig. 1}, nodes 1, 3 and 8 have highest degree, node 8 has highest betweenness centrality, also both nodes 1 and 8 have highest closeness centrality. Removing nodes 1 and 8 together would not create better effect of immunization than targeting node 8 alone. However taking nodes 1 and 9 together would fragment the network significantly. This ensemble issue termed as group centrality, has been introduced by Everett and Borgatti\cite{Everett_1999}. It explains the optimal solution for key-players problems. The group based centrality problem refers to the fact that selecting k nodes ensemble in a group is optimally better than selecting them individually. There are two major challenges related to the issue of finding key nodes for network immunization. First issue is to establish heuristic objective criteria; we need a heuristic approach to optimize objective criterion which challenge to achieve the target for identifying key nodes in the network aiming to the immunization for network vulnerability. Second issue is to resolve the group based centrality; the focus of group based centrality is to value the predetermined groups, however the requirement is to underline an approach to score individual nodes contributing to these alliances marginally, therefore a framework is desired to assign a solution to this problem. We target to address these stated issues with the objective of finding optimal solution for the node immunization problem in graphs. Motivated from the concept built upon cooperative game theory over networks we propose a solution to above stated problems. We aim to address the issue of defining objective criteria as game theoretic model for selection of key nodes in network based on their positional power and their functional influence over other nodes in the network. Given a constant k, we consider the problem of selecting k nodes which should be removed so that the spread of misinformation/virus over network. This heuristic takes care of objective criteria to model key players. We further propose a game theory based efficient algorithm which allows uniform ranking of individual nodes by computing its marginal contribution to create synergy in all possible coalitions of the nodes. The game theoretic efficient Shapley value approach defined by Michalak et al.\cite{Tomasz} provides a group based centrality framework for consistent ranking of individual nodes exploiting group results. This game theoretic framework incorporates the synergies of groups and works essentially the same way as regular centrality measures. In particular, for a given game where coalitions are allowed to configure, the fundamental issue is to assign the surplus acquired by the cooperation to all participating nodes as players. We propose to employ Shapley value\cite{Shapley}; a game theoretic scheme of dividing payoffs to the players. The Shapley value assigns a score to the player in individual as a function of its marginal contribution, allotted as the weighted average rise in payoff created by entry of the player. With this fundamental idea of game theoretic centrality approach on network, where nodes are the players, group of nodes are the coalitions and the payoffs of these coalitions are resolved to meet above stated preconditions. In this paper, we present the notion of finding Shapley Value based Information Delimiters (SVID) and introduce an efficient algorithm to calculate it. We propose a cooperative game that models the approach of defining objective criteria for information delimiters. We follow the work by Michalak et al.\cite{Tomasz} to undertake efficient computation of Shapley value using probabilistic approach upon nodes positive marginal contribution to the coalitions and model a polynomial running time algorithm for NP-Hard Shapley value computation\cite{Deng_1994}. We further propose an adaptive version of efficient  value algorithm, based on work by Adamczewski et al.~\cite{Adamczewski:2014:GSV:3006652.3006814} as Discount Shapley Value (DSV) to recompute the Shapley value after removing top ranked nodes which get immunized. We implement the proposed notion of SVID on real networks to detect network vulnerability in terms of average giant component size and robustness of the network in immunization procedure, when compared with benchmark algorithms. Finally, we evaluate the immunization strategy considering SIR epidemic.\subparagraph{}The rest of this paper is organized as follows. Section \ref{2} reports related studies. Section ref{3} provides preliminaries and  linked definitions. Section \ref{4} points up Shapley value based centrality and its efficient computation. Our proposed work is explained in Section \ref{5}. Section \ref{6} describes evaluation norms of simulated results. Section \ref{7} reviews performance of our proposed method using real and synthetic data sets and Section \ref{8} concludes the paper.
\section{Related Work}\label{2}
Based on structural centrality: There has been an important trace of work based on the nodes centrality subject to their importance in the network. Yingluo et al.~\cite{yingluo1993core} defined a node or a set of nodes to be vital when it contributes to reducing the size of giant connected component upon its removal. The shortest distance based methods which measure the loss to network connectivity rely on path-based-centralities like betweenness and closeness~\cite{Dolev_2010,hanneman2005introduction}. Dangalchev~\cite{Dangalchev_2006} introduced a new measure based on residual closeness centrality to define vulnerability of network when important vertices are removed. Another related issue is the importance of the bridging nodes among modules in the networks~\cite{Liu_2016,Hwang:2008:BCG:1401890.1401934,Chakraborty_2016, Narayanam:2014:SVA:3006652.3006762}. Similarly, other important work related to invulnerability research considered a node to be important if network becomes more agglomerate after its removal~\cite{Liu_2014}. Although, these closely related works based on different goals are prominent in related issues. However,for immunization goal they do not set up an optimal solution. Whereas, the selected criteria for SVID is designed for the target immunization, therefore it shows optimal results. Based on Eigen-spectra of network: First eigenvalue of graph adjacency matrix determines the vulnerability of the graph~\cite{Hong_2011,Chauhan_2009}. On this steam~\cite{Hong_2011,Chauhan_2009,Chen_2016} are the leading works based on graph eigen-spectra for the related issue of community detection. Based on influence maximization: Network node immunization and influence maximization are related problems with the same goal of affecting spread. The immunization algorithm works for immunizing a set of structural nodes in order to downturn the spread of the information/ virus in the network. On the other hand, objective for influence maximization is to achieve maximum influence spread in the network. Morone and Makse ~\cite{Morone_2015} suggested the similar concept of collective influence to find critical nodes in the network in order to dismantle the network efficiently and argue on the same line of relatedness of these problems. Subsequently, the size of largest connected cluster is a natural measure of influence when critical nodes are removed from the network~\cite{Moreno_2002}. Kempe et al.~\cite{Kempe_2003} identified influence maximization problem under operational dynamical process models (linear threshold, independent cascade), while Zhang et al.~\cite{Zhang_2016} considered different dynamical process model (SIR) for the same. Narayanam and Narahari~\cite{Narayanam_2011} proposed game theory based non scalable algorithm for large graphs. The pioneer works are either extension of the greedy algorithm or work on heuristics include~\cite{Goyal_2011,Chen_2010}. Network dynamics process: A related topic is to find vital nodes taking into account the target dynamic process in the network to amplify or downplay the spread. This approach receives dynamic influence approach relied upon various representative dynamics such as: rumor spreading model~\cite{Singh_2012}, voter model~\cite{Acebr_n_2005}, Kuramato model~\cite{Acebr_n_2005}, Ising model~\cite{Kempe_2003} and so on. Similarly, traffic dynamics is another application of dynamics of information networks for the control and the study of traffic and routing. Build upon traffic dynamics, Jiang and Liang~\cite{JIANG_2012} presented an approach to detect central nodes in traffic routes network efficient redistribute the congestion due to heavy traffic. Based on dynamics of disease spreading, Liu et al.~\cite{Liu_2014} gave measures to quantify epidemic spreading of nodes to counter disease spreading. Group centrality and other graph mining related approaches: The concept of group centrality applied to group of individuals and the groups with high centrality act as key players, Ortiz-Arroyo~\cite{Ortiz_Arroyo_2009} proposed a method to detect the set of key players under group centrality, if amputated can produce largest change in network connectivity entropy. Chen et al.~\cite{Chen_2010} defined shield value score for set of nodes for effective immunization strategy in large graphs. An analogous study of minimum dominating set (MDS) immunization is another related domain. This line of research has demonstrated and efficient MDS immunization in security vulnerability, traffic reduction in wireless sensor network~\cite{Yang_2015}, and contagion inoculation in social contact networks~\cite{Thakare_2016}. Another related study is the concept of structural hole which defines a position that can span and bridge diverge groups. Structural hole \cite{Lou_2013} can accelerate the spread in the network when targeted for dissemination, it can disconnect the network when removed if targeted for immunization~\cite{kang2011beyond}. Kleinberg et al.~\cite{Kleinberg_2008} proposed game based model that depends on bridging benefits as payoffs to the structural hole. Furthermore, Narayanam et al.~\cite{Narayanam:2014:SVA:3006652.3006762} proposed Shapley Value payoff division scheme based on defined characteristic function for gatekeeper locations to detect such topological nodes. Xu et al.~\cite{Xu_2017} used the notion of articulation nodes in graph to find structural hole spanners based on objective function of maximizing sum of all-pairs shortest lengths when they get eliminated from the graph.

\section{Preliminaries}\label{3} This section introduces the basic concepts and definitions from graph theory and cooperative (coalition) game theory for the network.\subsection{\textit{Definition 1(Graph):}}\textit{An undirected, unweighted network can be described by a graph $G(V,E)$, consists of nodes and edges. The sets of nodes (vertices) and edges are designated by $V(G)$ and $E(G)$ respectively. Nodes $u,v \in V(G)$, connected by one edge distance are called neighboring nodes and this edge is labeled by $(u, v)$. The neighboring nodes of a node $v$ are denoted by $N_v(G)$ and the degree of the node v is denoted as $degree_v(G)$.}\\Non-cooperative and cooperatives games are two broad categorizations of the games. In the non-cooperative games every player decides its action independently, whereas in cooperative games the player contributes profit to the coalition. We formalize the concept of cooperative games which are also known as coalition games.
\subsection{Definition 2 (Coalition game):}\textit{The coalition game is defined on $N=\{p_1,.....,p_{|N|}\}$be the set of players participating in a coalition game, a characteristic function $\vartheta:2^{V(G)}\rightarrow \mathbb{R}$ which depends on the graph $G$ and credits the share of contribution (in terms of real number payoffs) to the coalition $C \subset N$, where $ \vartheta (\emptyset ) = 0 $. Formally, a coalition game for the characteristic function $ \vartheta $ and coalition $N = V (G)$ is given by a tuple $< N, \vartheta >$}. The value of coalition is represented by $\vartheta(C)$. In case of coalition game on a network the players are nodes of the network. We use network $G$ for the coalition $ (V(G), \vartheta) $, where $V (G) = N$ represents the set of players and defines the characteristic function of this game. The grand coalition has all the players participating in the coalition and also carries highest value. The key investigation in coalition game theory is to assign the dividend from the coalition among all players in accordance to some criteria. In particular, Shapley~\cite{Shapley}  introduces the criteria for the concept of marginal contribution, a unique division scheme of computing weighted average payoff for a player that contributes to all possible coalitions it belongs to. The Shapley value confirms following preferred criteria:
\begin{enumerate}
\item \textit{efficiency}: the value acquired by the grand coalition due to contribution of all players is distributed among all of them.
\item \textit{symmetry}: if two players are symmetric in their roles they play to any coalition, their payoff is also symmetric.
\item \textit{null player}: a player with no contribution to the coalition receives null payoff.
\item \textit{additivity}: values of the two independent games add up to the worth for the sum of these games while together.
\end{enumerate}
Given a player $p_i \in N$ and a coalition $C\subset N$ where $p_i\notin C$, the marginal contribution of the player $p_i$ to the coalition $C$ is presented as $( \vartheta (C\cup{p_i})- \vartheta (C),\forall C\subset N)$. To define the notion formally, let $\pi \prod (N)$ be a permutation of players in $N$. Suppose players join the grand coalition one by one where all the sequences are equally likely. If $\pi(j)$ defines the position of $p_{j}$ in $\pi$ and let $C_{p_i}(i)$ be a coalition formed by all the players who come before the player $p_i$ in permutation $\pi$, then $C_{\pi}(i)$ can be represented as $C_{\pi}(i) = \{p_i \in \pi: \pi(j) < \pi(i)\}$.
\subsection{Definition 3 (Shapley value):}\textit{The Shapley value of player $p_i$ denoted as $\Theta(p_i)$ is characterized as the average marginal contribution that $p_i$ made to coalition $C_\pi(i)$ over all $\pi \in \prod$}:
\begin{equation}
\Theta(p_i)=\frac{1}{\left |N \right |!}\sum_{\pi\in \prod}[ \vartheta (C_{\pi }(i)\cup\left \{ p_{i} \right \})- \vartheta (C_{\pi }(i))]\label{eq:1}
\end{equation}
The above equation can be simplified to the equivalent one:
\begin{equation}
\Theta(p_i)=\sum_{C\subseteq N\setminus \{p_i\}}\frac{| C |!(|N|-|C|-1)!}{|N|!}[ \vartheta (C\cup\{p_i\})- \vartheta (C)]\label{eq:2}
\end{equation}
\section{Shapley value based centrality and its efficient computation}\label{4} From Shapley value calculation as shown in Eq.2, it is readily apparent that it involves $N!$ number of permutations to find marginal contribution of a player, hence the computational expense of direct implication of the game is very high. Narayanam and Narahari~\cite{Narayanam_2011} proposed SPIN algorithm, a Monte Carlo simulation approach to find influential nodes based on Shapely value approach for target set selection problem. Furthermore, Michalak et al.~\cite{Tomasz} considered a probabilistic approach to deal with sampling in Monte Carlo simulation to find exact solution for the same game to compute Shapley value in polynomial time. On the same line, Adamczewski et al.~\cite{Adamczewski:2014:GSV:3006652.3006814} proposed discount Shapley value approach to recompute the Shapley values in the remaining graph each time nodes are selected as targets for spread/immunization. We introduce all these related concepts in this section.
\subsection{Shapley value approach for centrality defined by SPIN algorithm:}In the context of information dissemination, Narayanam and Narahari~\cite{Narayanam_2011} defined Shapley value based solution concept to find  influential nodes based on the spread impact of a node on neighboring nodes and defined its characteristic function $ \vartheta : 2^{V (G)}\rightarrow \mathbb{R}$ as the size of one hop neighboring node set, formally defined as:
\begin{equation}
\vartheta (C)=\left\{\begin{matrix}
0 &if \, C=\varnothing   \\ 
\left | neighbors(C) \right | &otherwise 
\end{matrix}\right.\label{eq:3}
\end{equation}
\subsection{Exact solution for efficient computation of SPIN}Considering the same game defined in Eq.3, Michalak et al.~\cite{Tomasz} offered polynomial time $O(| V (G)| + | E(G)|)$ exact solution for computing Shapley value. Authors present the characteristic function for the same as:
\begin{equation}
\vartheta (C)=\left\{\begin{matrix}
0 &if \, C=\varnothing   \\ 
\left | fringe(C) \right | &otherwise 
\end{matrix}\right.\label{eq:4}
\end{equation}
\hspace*{3ex}The fringe defined in the characteristic function meant by the set of all nodes in coalition and their neighborhood sphere by an edge distance, therefore both defined characteristic functions deliver the same meaning for Shapley value calculation of the node. Authors explained the probabilistic approach to appraise Shapley value of a node $v$ by considering all permutation where $v$ makes positive contribution on its inclusion to the coalition and bypassing other non-trivial permutations when there is void contribution by the node. The investigation of necessary and sufficient condition for a node $v$ to fringe the coalition $C$ marginally, can be formulated as the probability that no neighbor $u \in N_v(G)$ of the node $v$, or no edge $(u,v)$ enters the coalition before node $v$ does. The solution for such status holds true with probability: $1/1+degree_v(G)$. The closed formula for the calculation of Shapley value of node $v$ for the same game as per Eq.~\eqref{eq:4} is given by:
\begin{equation}
\Theta( v )=\sum_{u\in v\cup N_v(G)}\frac{1}{1+deg_u(G)}
\end{equation}
\subsection{Discounted Shapley Value (DSV):}An adaptive version of the approach for exact solution of Shapley value considers the influence  of the node $v$ on the other nodes only exclusive of the effect onto itself. DSV algorithm~\cite{Adamczewski:2014:GSV:3006652.3006814} also explains to remove the neighboring nodes of selected top nodes for target set as best influencers and subtract the contributions of these neighboring nodes to the Shapley values common neighbors of top nodes and their one hop neighboring nodes.
\section{Proposed Shapley Value based Information Delimiters (SVID)}\label{5} In this section we propose our approach for the information delimitation problem. We adopt the same framework as reviewed in the efficient computation of SPIN game in the previous section and consider the formation of game on undirected and unweighted network. The problem of information delimiters works in the scenario of immunization, however it is converse the problem of information propagation as dealt by game defined in SPIN. The insight behind formation of objective criteria for the game targeting SVID aims at considering those players (nodes) whose removal would either increase short term distances among rest of the nodes or decrease number of alternative paths among them. The norm of number of common neighbors up to h-hops apparently translates the norm of having
number of alternative paths. Fewer the common neighborhood, limited the possibility of having alternative paths among nodes in the network. Increase in the distance between nodes would fairly lead to information die out quickly. This rationale behind above described paradigm affirms the heuristic objective criteria for SVID and Shapley value based solution for this problem involves group synergies to consider group based centrality precondition.
\subparagraph{}For a coalition $C$, a node $u$ can marginally contribute a neighboring node $v$ in two cases: either both are present in different connected components of the graph and have no common neighbor, or both are connected by an edge and have some common neighbors. Since in the former case marginal contribution is 1, for the later case we examine the nodes $u$ and $v$ connected by an edge and having $K$ common neighbors up to one hop to determine the random probability that a node $u$ can positively contribute a neighbor $v$, in coalition $C$. Considering common neighbors up to two levels, can perceive all paths between two nodes up to length four. Since, Shapley value calculation is based on enumerating all the possible permutations; therefore we calculate the probability of desired permutations in which a node $u$ can contribute its neighbor $v$ to the coalition. The requirement is that the node $u$ must join the coalition before node $v$ and all the common neighbors between $u$ and $v$ must join the coalition after these two nodes, it is possible that $v$ can be attributed to the contribution of these common neighbors as node $u$ already present in the coalition. Suppose there are $K$ common neighbors that exist between two nodes, then the explanation of our proposed method goes as follows:
\begin{enumerate}
\item\textit{Let us select $K + 2$ positions in the ordering of all elements from N. We can do such selection in $^{N}\textrm{C}_{K+2}$ number of ways}.
\item \textit{In the last $K$ picked positions, bring in all elements from common neighbor list . Precisely before these elements, place the elements $u$ and $v$, the nodes under the consideration. The number of such selections is $K!$}.
\item \textit{The remaining components can be positioned in $(|N| - (K + 2))!$ different ways}.
\item \textit{Therefore, the total number of such permutations are}:
\begin{equation*}
^{N}\textrm{C}_{K+2}*(K!)*(|N| -(K+2))!= \frac{|N|!}{(K+1)(K+2)}
\end{equation*}
\item The probability for random selection of one of such permutations is:
\begin{equation*}
\frac{1}{(K+1)(K+2)}
\end{equation*}
\end{enumerate}
Next, We present SVID algorithm in adaptive version and regulate not to examine a node for the aim of information delimitation if its neighbor has been chosen for the same because this may lead to the removal of nodes from a particular portion of the graph and the rest of graph will remain active for information spreading. Also, after the removal of some percentage of the nodes, the graph topology changes significantly, therefore we recalculate the Shapley values after the removal of top-k nodes from the graph as proposed by Adamczewski et al.~\cite{Adamczewski:2014:GSV:3006652.3006814}. The Shapley value formulated in the proposed algorithm (Algorithm 1) takes care of the articulation points of the graph who are benefited by location being information spreaders between the components and also of the hubs who are connecting large groups of immediate neighbors. As the number of common neighbors between the neighbors of bridge node would be zero, $1/(K + 1)(K + 2)$ takes its maximum value. Also, for the hubs having greater number of neighbors, the summation over all the neighbors takes care for their contribution to this score. For each edge in the graph in the worst case, we can traverse the complete graph in the breadth first fashion, therefore Time Complexity = $O(|E|(|V | + |E| + |V|)) = O(|E|^2)$. As ordered by proposed algorithm and related to the size of the network under consideration, top nodes are removed from the network. Next section covers experimental simulation of the same on synthetic and real-world data sets.
\begin{algorithm*}[t]
\DontPrintSemicolon
\KwIn{Unweighted, undirected graph $G(V,E)$}
\KwOut{Shapley values of all nodes}
\For {each edge $e\in E(G)$}
{Remove the edge $e$ connecting $u$ and $v$ from $G$\;
$S_v\leftarrow$ Nodes reachable from $v$ upto one hop\;
$S_u\leftarrow$ Nodes reachable from $u$ upto one hop\;
$K\leftarrow|S_v \cap S_u |$\;
$\Theta(u)+=\frac{1}{(K+1)(K+2)}$\;
$\Theta(v)+=\frac{1}{(K+1)(K+2)}$\;
}
delimiters $\leftarrow0$; A$\leftarrow0$; $k\leftarrow$ no. of top nodes\;

\For {1 to k}
{
\If{not all nodes are immunized}
 {top nodes $\leftarrow argmax_{v\notin delimiters}\{\Theta(v)\}$\;
  delimiters $\leftarrow delimiters \cup \{topnodes\}$\;
  A$\leftarrow$A$\cup\{topnodes\}$\;
  \For {each $u\in N_{topnode}(G)$}
   {
    $v\leftarrow N_u (G)$\;
    \For {each edge(u,v) and K$\leftarrow |S_u \cap S_v|$}
    { $\Theta(v)- =\frac{1}{(K+1)(K+2)}$\;}
    }
   }  
\Else
 {select a node $i\notin$ A with highest $\Theta(i)$ and add to A\;
 }
}
\Return{(A containing top k delimiters)}

\caption{SVID Adaptive algorithm}\label{algo:svida}
\end{algorithm*}
\section{Numerical simulation setup} \label{6} In this section, we define experimental setup to evaluate the efficacy of our proposed approach across other benchmark approaches. With this objective in mind, we consider two model networks and four real-world networks (\ref{Table 1}) for simulation. We examine the performance of our proposed algorithm on the basis of: network vulnerability, robustness and epidemic spreading through SIR model (Algorithm 2). \textbf{Network Vulnerability:} The Minimum size of largest connected cluster(lcc) of the network comes up as a result of immunization procedure, which determines the better performance of approach under consideration. \textbf{Robustness:} Robustness~\cite{Schneider_2011} quantifies the performance of immunization process by considering the size of lcc at every $q$ fraction of nodes are removed from network. It keeps account of all possible damage in network even before total collapse. Robustness $(R)$ is defined by as:
\begin{equation}
R=\frac{1}{N}\sum_{Q=1}^{N}s(Q)
\label{eq:6}
\end{equation}
where $s(Q)$ is the fraction of lcc size after $Q = qN$ nodes are removed from the network of size $N$, at each step. Precisely, the smaller the $R$, better the immunization method is. \textbf{Epidemic Spreading:} SIR epidemic spreading model is the most explored spreading criteria for complex network. We evaluate the proposed algorithm by dynamic simulation of infection propagation on the observed network. In this model nodes carry one of the states being susceptible $s$, infected $i$ or recovered $r$. The fraction of infected nodes decrease with effective portion of delimiters /immunized nodes in the network having infectious nodes present. We also find the final absolute portion of the recovered nodes at steady state when no more infected nodes are present in the network as a performance benchmark for different immunizing strategies.
\subsection{SIR algorithm} To evaluate the effect of spreading, we consider the classic SIR model defined in~\cite{Gupta_2016}, as described in Algorithm 2. All nodes present in the network are partitioned in sets of three states described as susceptible, infected, and recovered. A node in network can remain in one of these states. These states change as epidemic progress and can be described as functions of time $t$: $s(t), i(t)$ and $r(t)$. Each time step an infected node is chosen and it can infect its immediate susceptible neighbors with probability $\lambda$. An infected node can change its state to recovered with probability $\sigma$. The recovered nodes are immune to subsequent infection and do not change their state. The spreading dynamic attains steady state when no more infected nodes are left. Considering the heterogeneity of complex network due to presence of nodes with varying degree and association within the network, we analyze time evolution density for susceptible, infected and recovered nodes having degree $k$ as $s_{k}(t)$, $i_{k}(t)$, and $r_{k}(t)$. Where $s_{k}(t)+i_{k}(t)+r_{k}(t)=1$ holds true for $k$ at time $t$. The SIR model on complex network is considered to have mean degree $\left \langle k \right \rangle$ and degree distribution $P(k)$, where $\left \langle k \right \rangle=\sum_{k}kP(k)$. The mean-field approach for dynamics of contagion states in the network describes realization of Markov chain interaction with following set of differential equations.
\begin{align}
&\frac{\mathrm {d}s_{k}(t) }{\mathrm{d} t}=-\lambda k s_{k}(t)\omega(t)\\
&\frac{\mathrm{d}i_{k}(t) }{\mathrm{d} t}=\lambda k s_{k}(t)\omega(t)-\sigma i_{k}(t)\\
&\frac{\mathrm{d}r_{k}(t) }{\mathrm{d} t}=\sigma i_{k}(t)
\end{align}
 Where $\omega(t)=\sum_{k'}i_{k}(t)P(k/k')$. Here network is considered to have no degree-degree correlation and $P(k/{k}')$ is the  conditional probability of connecting a node having degree ${k}'$ to the node having degree $k$. The conditional probability $P(k/{k}')$ is proportional to $kP(k)$ in the case of uncorrelated network. The epidemic threshold $\lambda_{c}=\left \langle k \right \rangle/\left \langle k^{2} \right \rangle$ for finite size  regular networks where$\left \langle k \right \rangle<\infty$, has a finite value. Whereas, in case of scale free networks $\lambda_{c}$ declines to die out as network size $N$ increase(i.e.$\left \langle k \right \rangle\rightarrow\infty$ hence $\lambda_{c}\rightarrow 0$). In that event   $\lambda_{c}$ is a trivial property for the large size of networks.
\begin{algorithm}
\DontPrintSemicolon
\KwIn{A graph with initial susceptible population,$\lambda$, and $\sigma$ }
\KwOut{The number of infected population $i(t)$ during the process at time $t$}
Choose $i$ number of nodes at random and add them to infected list\;
\While{size of infected list >0}
{
\For{each node u in infected list}
 {\If {$v\in N_u(G)$ is susceptible}
  {with probability $\lambda$, move $v$ to infected list\;
   increase $t$ by $1$\;
  }
  \Else{\If {$v \in N_u(G)$ is infected}
    {with probability $\sigma$, move $v$ to recovered list
    }    
   }
 }
}
\Return{i(t)}

\caption{SIR Algorithm}\label{algo:sir}
\end{algorithm}
\subparagraph{}We evaluate our proposed algorithm considering two model networks, also known as Erd$\ddot{o}$s-R$\dot{e}$nyi(ER) and Scale-Free(SF). We construct ER network of 5000 nodes with $\left \langle k \right \rangle = 3.5$ and 10000 edges, Barab$\acute{a}$si-Albert SF network of 5000 nodes with power-law exponent $\gamma = 2.7$. We also consider four real-world networks (table~\ref{Table 1}). The empirical networks are from various fields: Autonomous System (AS)~\cite{nr} is an undirected network of connected internet protocol routing prefixes,where nodes represent autonomous system and edges denote connection. Email~\cite{Leskovec_2007} is a communication network, where nodes are users of unique email id and emails between users represent undirected edges. Internet~\cite{rossi2013topology} dataset is a technological network having nodes as hosts and edges as connections between hosts. Douban~\cite{reza} is an online social network providing recommendation reviews sharing services between users,where nodes are the users and edges are friendship among them. To validate the efficiency of proposed method (SVID), we compare it with benchmark target immunization strategies based on the degree centrality, the betweenness centrality, the eigenvector centrality and the coreness centrality. The top nodes with highest orders, according to these centrality criteria are defined as target nodes to be immunized. We use adaptive version of these strategies and recalculate the nodes importance after every $5\%$ nodes are selected as immunized. A node is selected as candidate for immunization, only if its neighbors are not considered for the same job. The adaptive variation of benchmark strategies are taken as degree adaptive (DA), betweenness adaptive (BWA), eigenvector adaptive (EVA) and coreness adaptive (CNA). We also consider an adaptive version of our proposed method SVID adaptive (SVIDA), as described in Algorithm 1.
\begin{table}
\caption{Basic statistical features of the real-world data sets including the numbers of nodes and edges, the maximum degree $k_{max}$ and the clustering coefficient $c$.}

\begin{tabularx}{\textwidth}{XXXXX}
  \hline
 Networks & Nodes & Edges & $k_{max}$ & $c$\\ \hline
AS & 6,476 &13,895 &1500 &0.252\\
Email &32,430 &54,393 &623 &0.1136 \\ 
Internet &40,164 &85,123 &3400 &0.205\\ 
Douban &1,54,908 &3,27,162 &287 &0.0160\\  
  \hline
\end{tabularx}
\label{Table 1}
\end{table}

\section{Results and discussion}\label{7} 
\begin{figure}[!t]
            \begin{subfigure}[t]{0.24\textwidth}
                \includegraphics[width=1\linewidth,height=3cm]{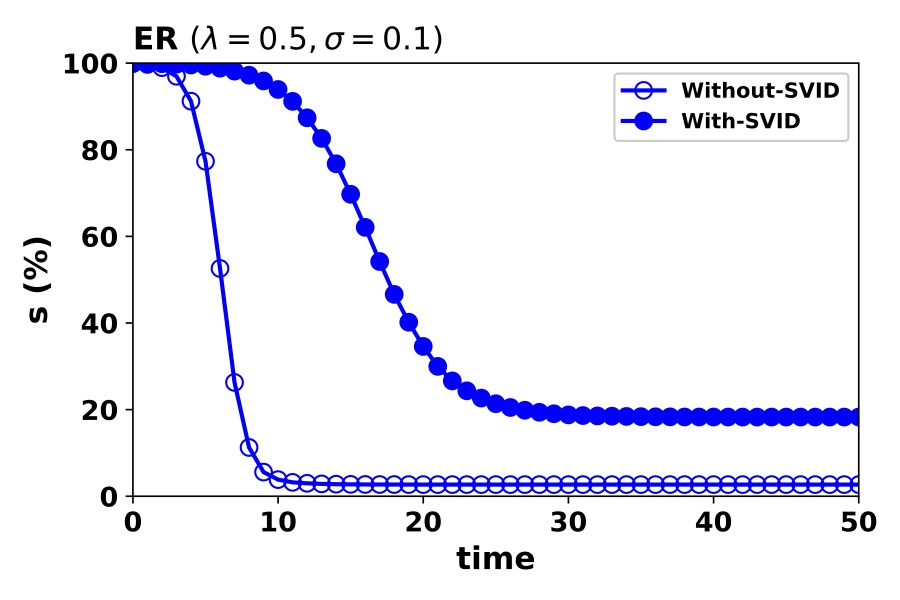}
                \caption{}
            \end{subfigure}
            \begin{subfigure}[t]{0.24\textwidth}
                \includegraphics[width=1\linewidth,height=3cm]{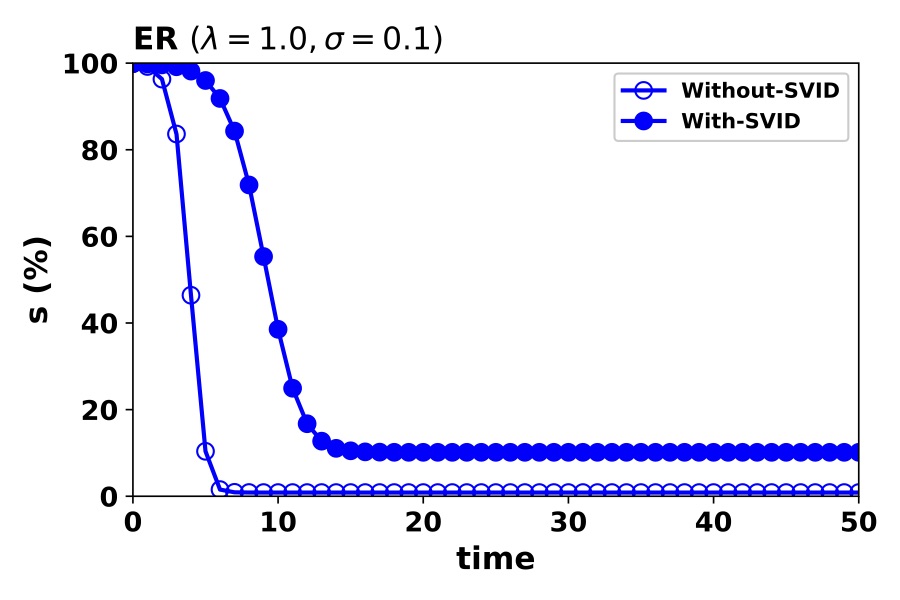}
                \caption{}
            \end{subfigure}
            \begin{subfigure}[t]{0.24\textwidth}
                \includegraphics[width=1\linewidth,height=3cm]{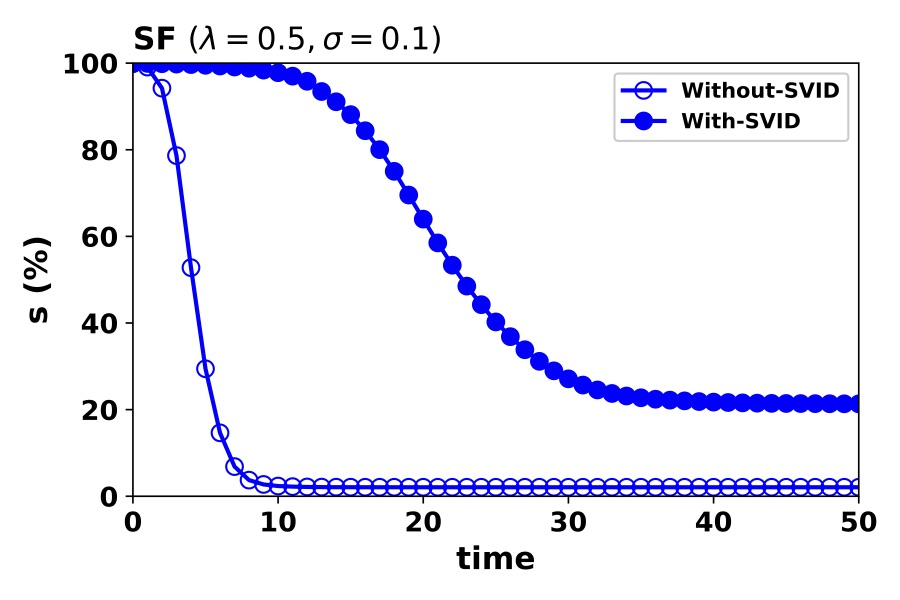}
                \caption{}
            \end{subfigure}
            \begin{subfigure}[t]{0.24\textwidth}
                \includegraphics[width=1\linewidth,height=3cm]{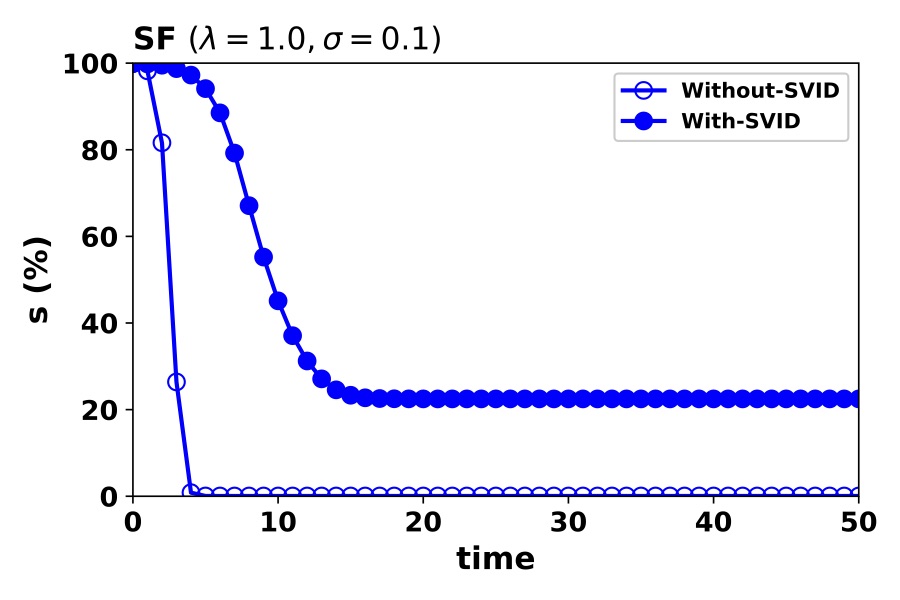}
                \caption{}
            \end{subfigure}
		
            \begin{subfigure}[t]{0.24\textwidth}
                \includegraphics[width=1\linewidth,height=3cm]{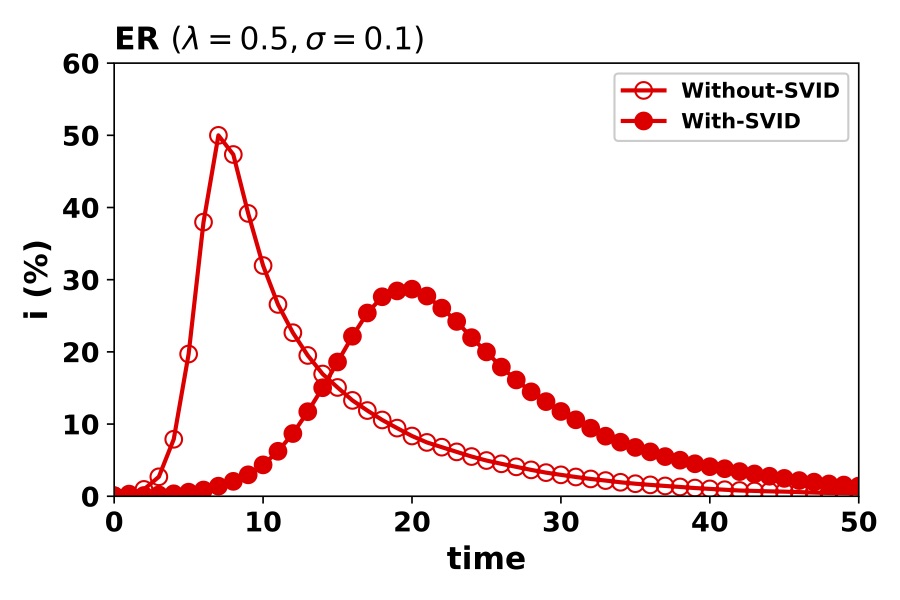}
                \caption{}
            \end{subfigure}
            \begin{subfigure}[t]{0.24\textwidth}
                \includegraphics[width=1\linewidth,height=3cm]{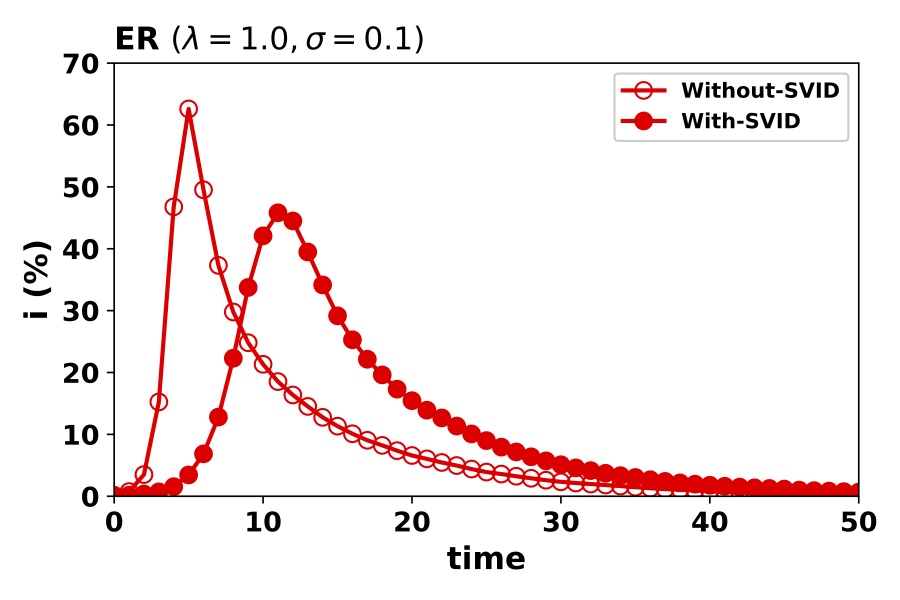}
                \caption{}
            \end{subfigure}
            \begin{subfigure}[t]{0.24\textwidth}
                \includegraphics[width=1\linewidth,height=3cm]{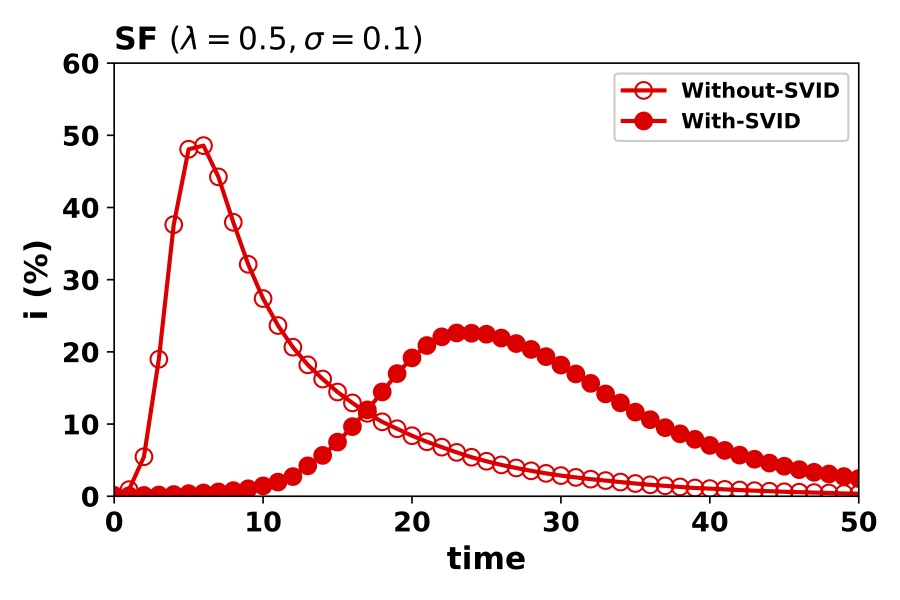}
                \caption{}
            \end{subfigure}
            \begin{subfigure}[t]{0.24\textwidth}
                \includegraphics[width=1\linewidth,height=3cm]{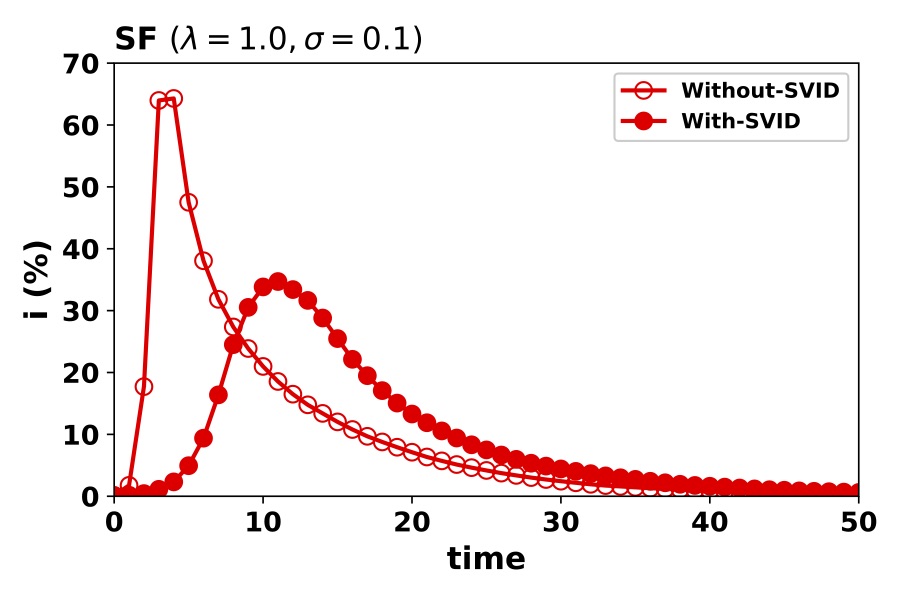}
                \caption{}
            \end{subfigure}
            
            \begin{subfigure}[t]{0.24\textwidth}
                \includegraphics[width=1\linewidth,height=3cm]{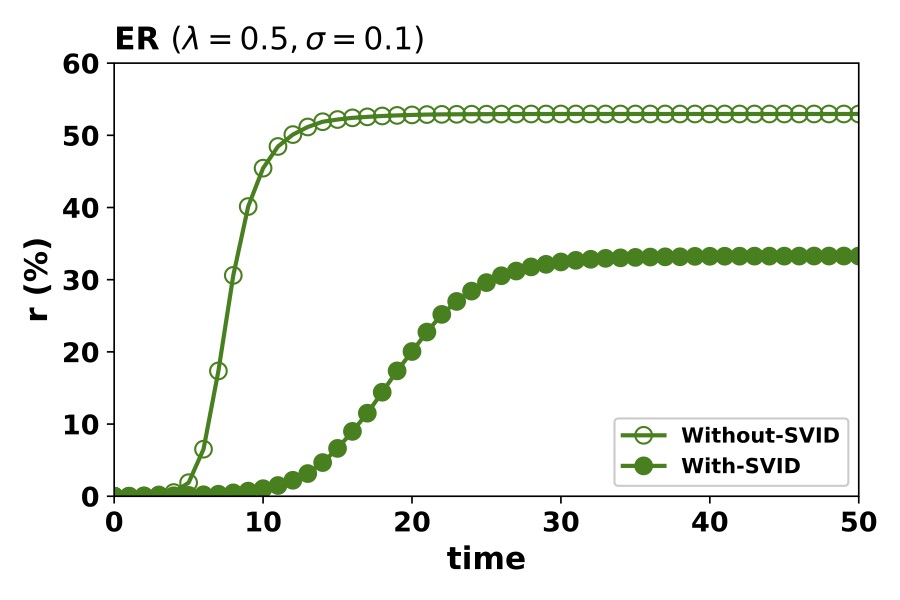}
                \caption{}
            \end{subfigure}
            \begin{subfigure}[t]{0.24\textwidth}
                \includegraphics[width=1\linewidth,height=3cm]{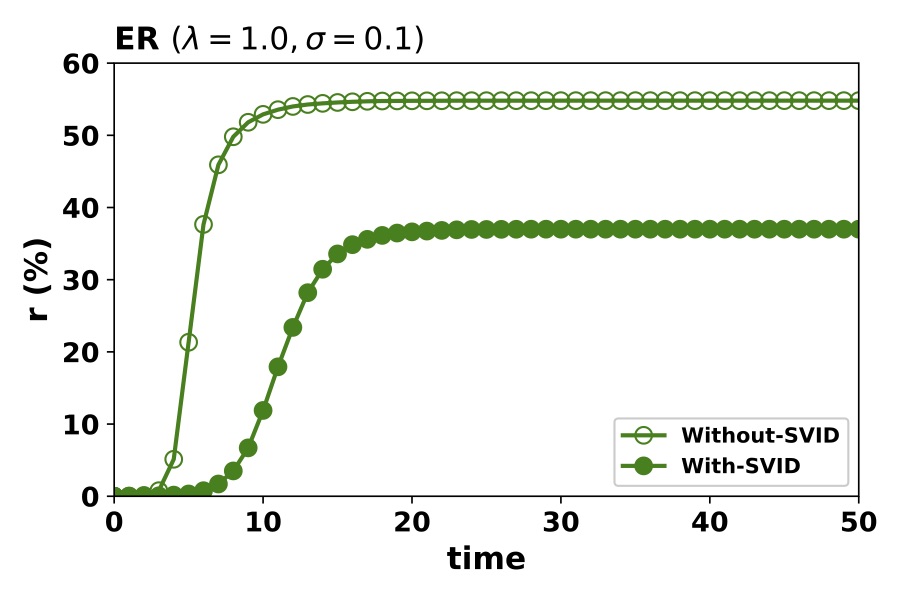}
                \caption{}
            \end{subfigure}
            \begin{subfigure}[t]{0.24\textwidth}
                \includegraphics[width=1\linewidth,height=3cm]{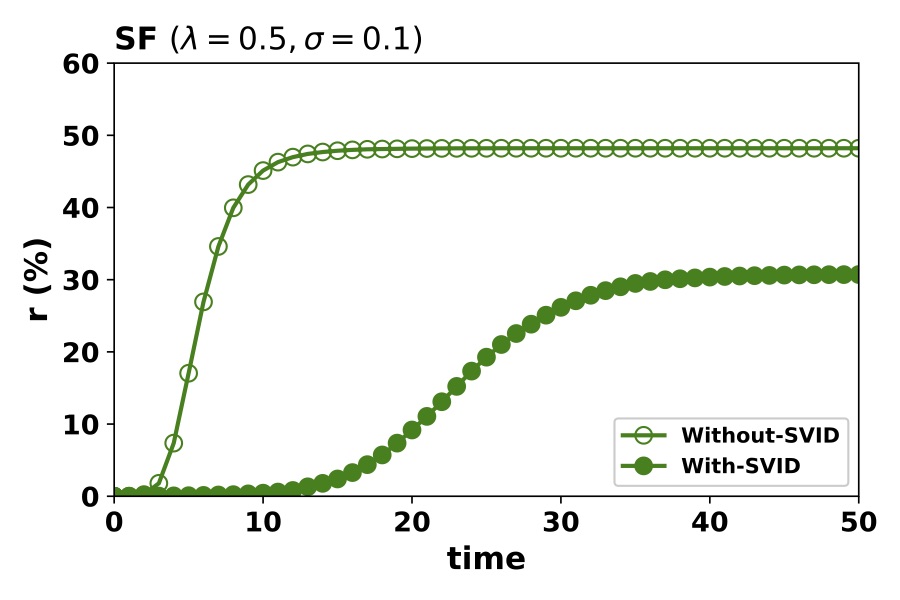}
                \caption{}
            \end{subfigure}
            \begin{subfigure}[t]{0.24\textwidth}
                \includegraphics[width=1\linewidth,height=3cm]{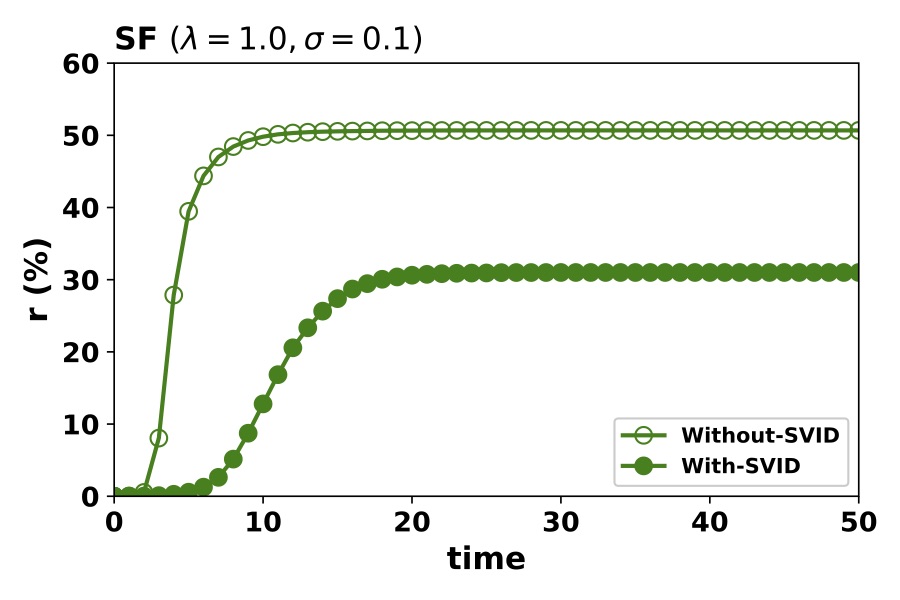}
                \caption{}
            \end{subfigure}
          
           \caption{Under different range of $\lambda$ and $\sigma$, susceptive s(\%), m/infected i(\%) and recovered r(\%) population against time for model networks(Erd$\ddot{o}$s-R$\dot{e}$nyi ER and Scale-Free SF ) without immunization and with immunization under SVIDA for $q=15\%$ fraction of immunized population.}
 \label{Fig. 2} 
\end{figure}

To predict the performance of proposed strategy, the SIR epidemic spreading model is used to analyze the immunization validness on the ER network and the SF network. In SIR simulation, each node belongs to one of the states from susceptible(s), infected(s) or recovered(r) for different values of infect rate $\lambda$ and recover rate $\sigma$. At $t=0$ all nodes are consider as susceptible and a node randomly chosen is set to be infected. We implement SIR model for network without delimiters and network with delimiters for immunized nodes $(q = 15\%)$ including incident edges are removed from the network selected according to SVIDA. The experiment is conducted for sufficient runs (50) and average of all executions is plotted for all the values against time. For both implementations, we compare dynamics in the population of susceptible, infected and recovered population fraction in percentage with time progression, as shown in \ref{Fig. 2}. We use $\lambda=1.0$ and $\sigma=0.1$ for the strong infection simulation and $\lambda=0.5$ and $\sigma=0.1$ for moderate infection simulation. For all the model networks, infected fraction $i(\%)$ is significantly lower when using $q$ fraction of delimiters as immunized population.\\ \hspace*{3ex} The simulation results for fraction of susceptible, infected and recovered population are shown column wise for different ranges of $\lambda$. We also investigate the absolute value for recovered nodes $|r|$ at steady state of the infection when there is no more infected nodes left in the network. To find $|r|$, we implement epidemic process on the network for suitably large value of time, as eventually the epidemic dies at time-step = $\infty$. We consider recover rate $\sigma= 0.1$ and infect rate $\lambda$ with values $0.2$ for low infection, $0.8$ for high infections. We implement sizable number runs for SIR epidemic process for all data sets and consider standard deviation of results by plotting errors with results $|r|$ vs $q$. The results of various immunizing strategies based on adaptive (degree, betweenness, eigenvector and coreness) compared with SVIDA for $|r|$ at different proportion of immunized nodes $(q)$ selected are shown in \ref{Fig. 3}.
\begin{figure}[!h!b]

            \begin{subfigure}[b]{0.49\textwidth}
                \includegraphics[width=1\linewidth,height=6.0cm]{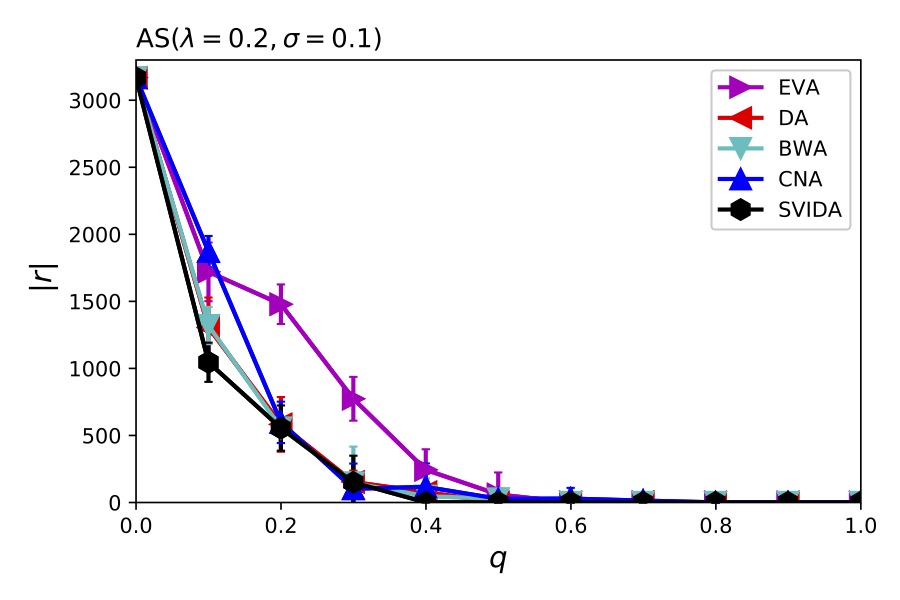}
                \caption{}
            \end{subfigure}
            \begin{subfigure}[b]{0.49\textwidth}
                \includegraphics[width=1\linewidth,height=6.0cm]{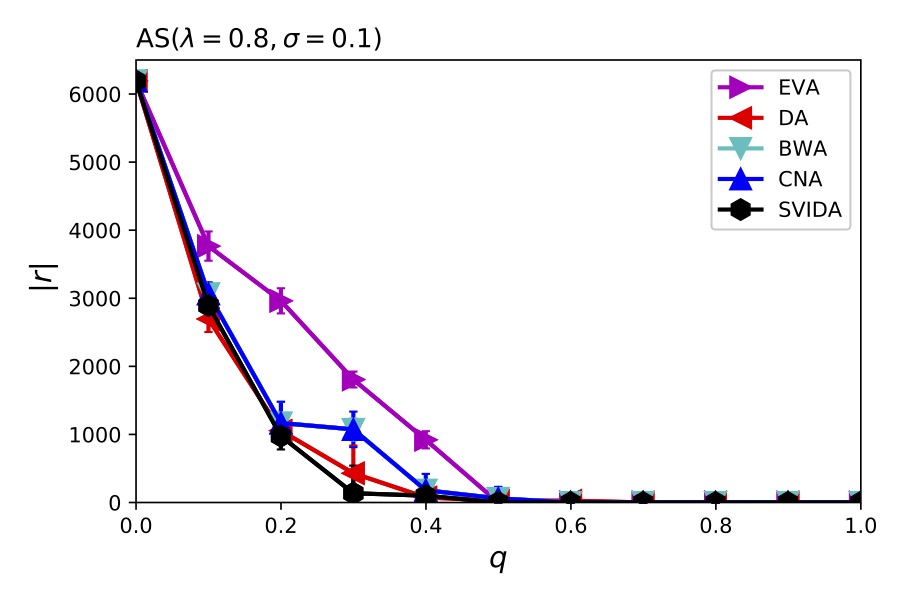}
                \caption{}
            \end{subfigure}
       
            \begin{subfigure}[b]{0.49\textwidth}
                \includegraphics[width=1\linewidth,height=6.0cm]{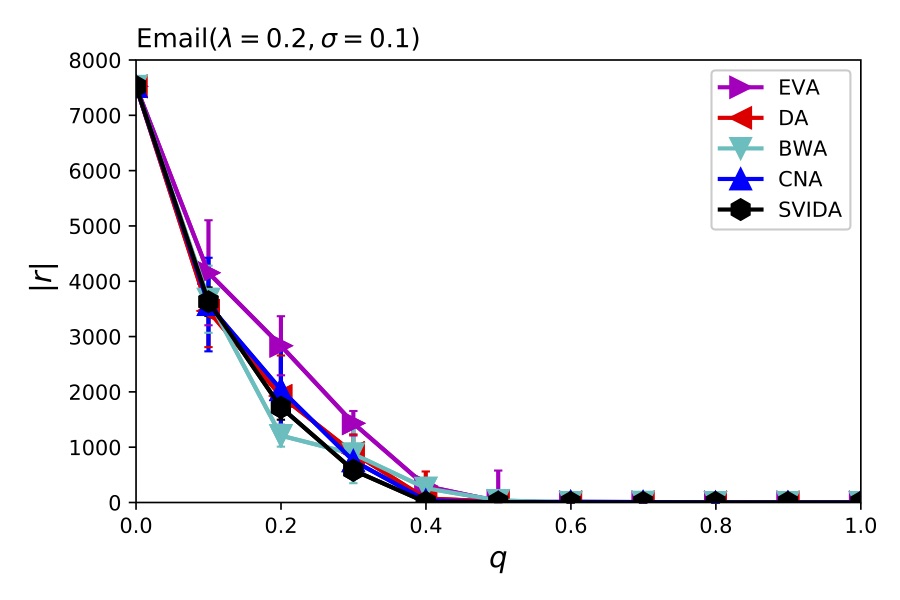}
                \caption{}
            \end{subfigure}
            \begin{subfigure}[b]{0.49\textwidth}
                \includegraphics[width=1\linewidth,height=6.0cm]{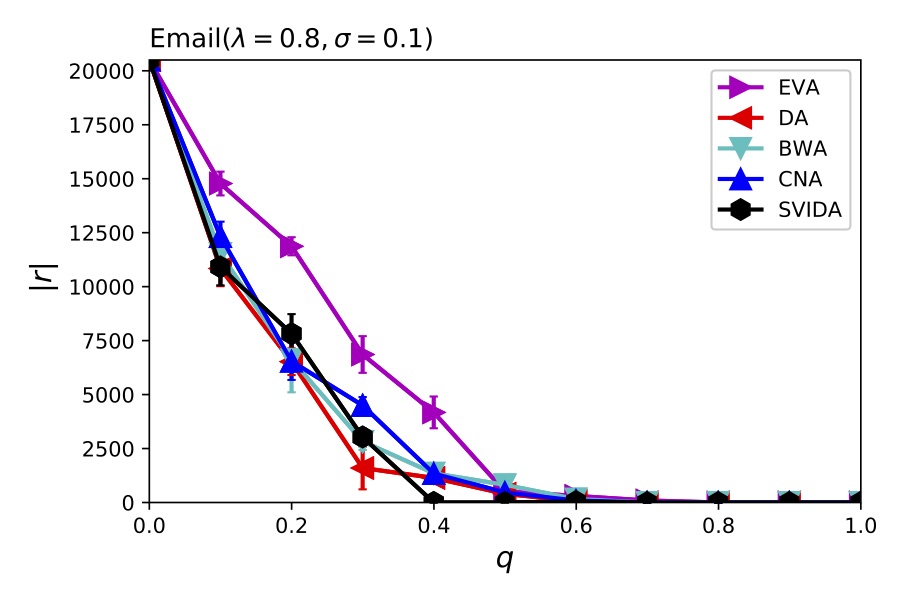}
                \caption{}
             \end{subfigure}
\end{figure}
 \begin{figure}[!t!h]\renewcommand\figurename{Fig.}
 \ContinuedFloat              
                
             \begin{subfigure}[]{0.49\textwidth}
                \includegraphics[width=1\linewidth,height=6.0cm]{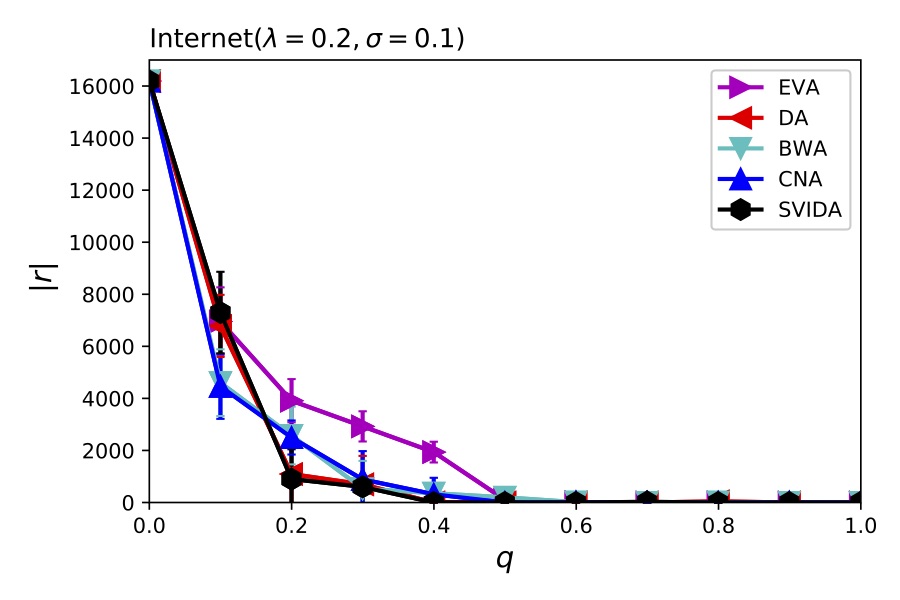}
                \caption{}
            \end{subfigure}
            \begin{subfigure}[]{0.49\textwidth}
                \includegraphics[width=1\linewidth,height=6.0cm]{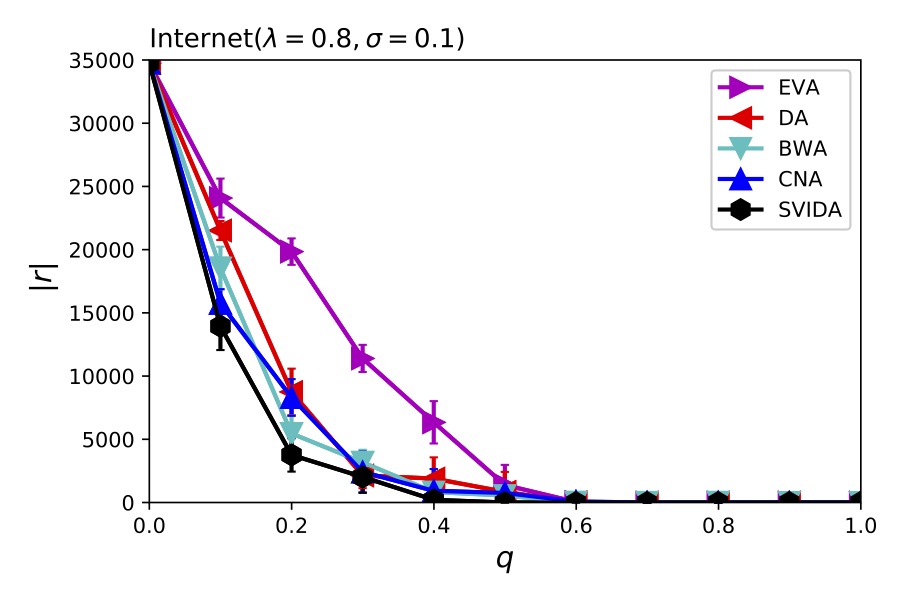}
                \caption{}
			\end{subfigure}
			
             \begin{subfigure}[]{0.49\textwidth}
                \includegraphics[width=1\linewidth,height=6.0cm]{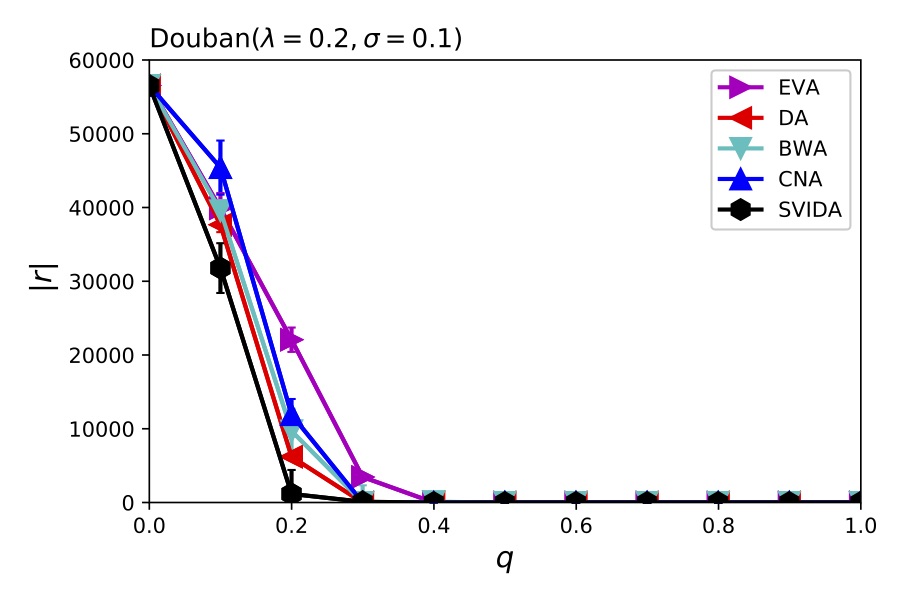}
                \caption{}
            \end{subfigure}
            \begin{subfigure}[]{0.49\textwidth}
                \includegraphics[width=1\linewidth,height=6.0cm]{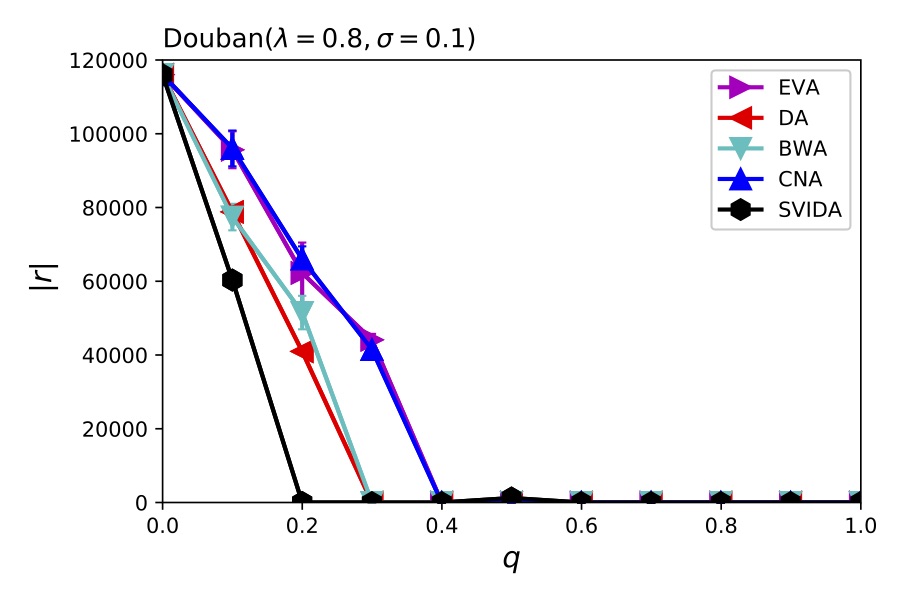}
                \caption{}        
            \end{subfigure}

           \caption{Absolute value of total recovered nodes $|r|$ at steady state of infection under various immunization strategies for range of immunization nodes fraction $q$ with varying infect rate $\lambda=0.2,0.8$ and recover rate $\sigma=0.1$.}
\label{Fig. 3}           
\end{figure}
\begin{figure}

			\begin{subfigure}[t]{0.49\textwidth}
                \includegraphics[width=1\linewidth,height=6cm]{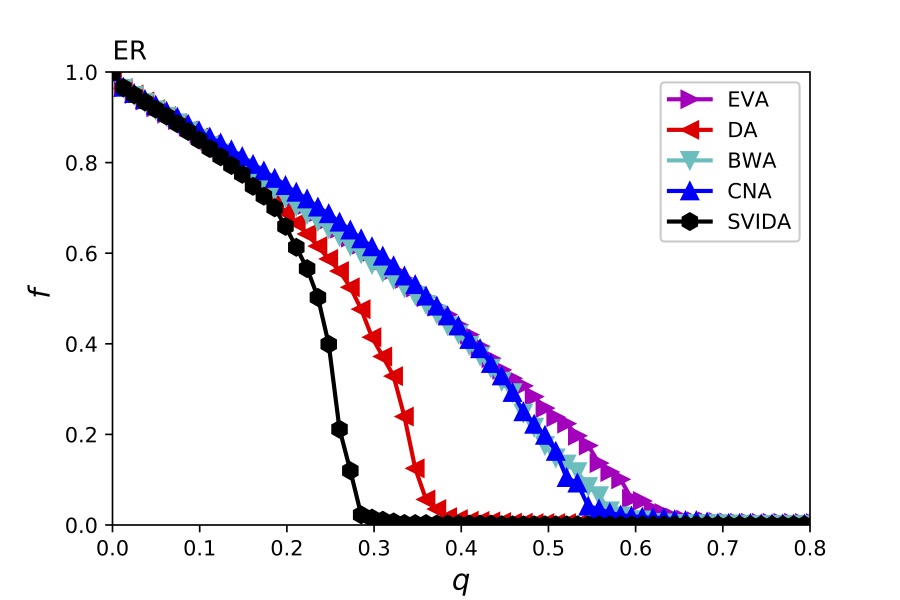}
                \caption{}
            \end{subfigure}
            \begin{subfigure}[t]{0.49\textwidth}
                \includegraphics[width=1\linewidth,height=6cm]{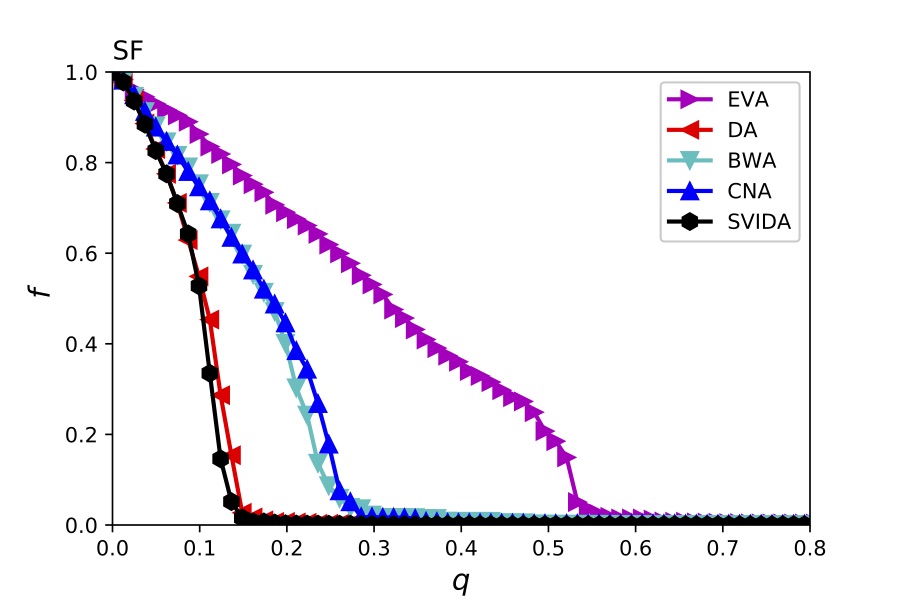}
                \caption{}
            \end{subfigure}
\caption{The fraction of lcc (largest connected cluster) $f$ vs immunized nodes fraction $q$ under various strategies for model networks (ER, SF).}\label{Fig. 4}
\end{figure}
\begin{figure}

            \begin{subfigure}{0.49\textwidth}
                \includegraphics[width=1\linewidth,height=6cm]{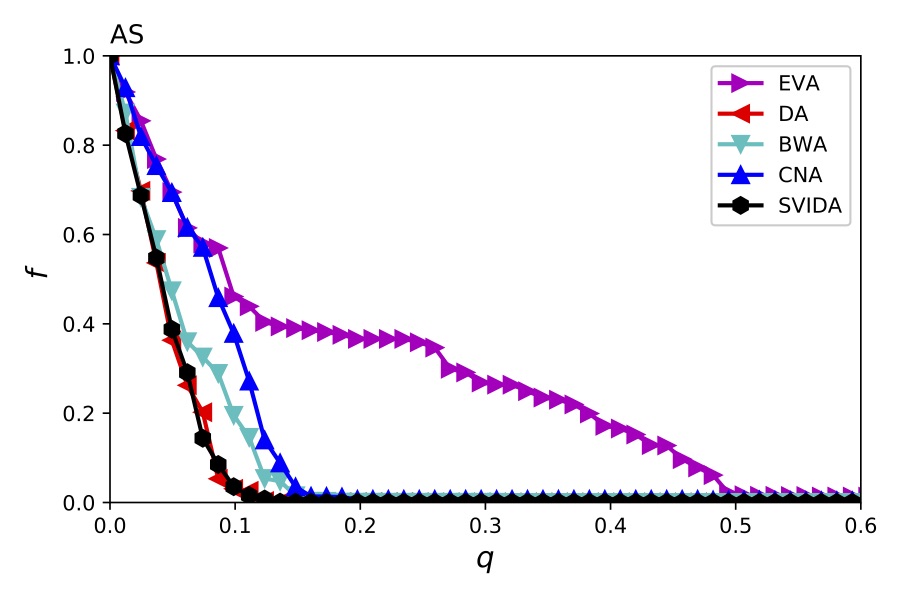}
                \caption{}
            \end{subfigure}
            \begin{subfigure}{0.49\textwidth}
                \includegraphics[width=1\linewidth,height=6cm]{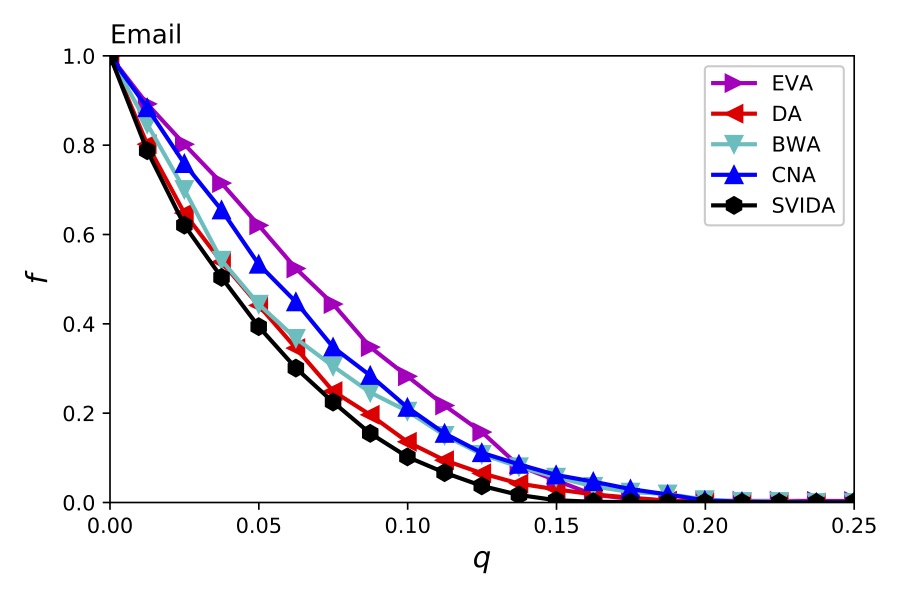}
                \caption{}
            \end{subfigure}
			           
            \begin{subfigure}{0.49\textwidth}
                \includegraphics[width=1\linewidth,height=6cm]{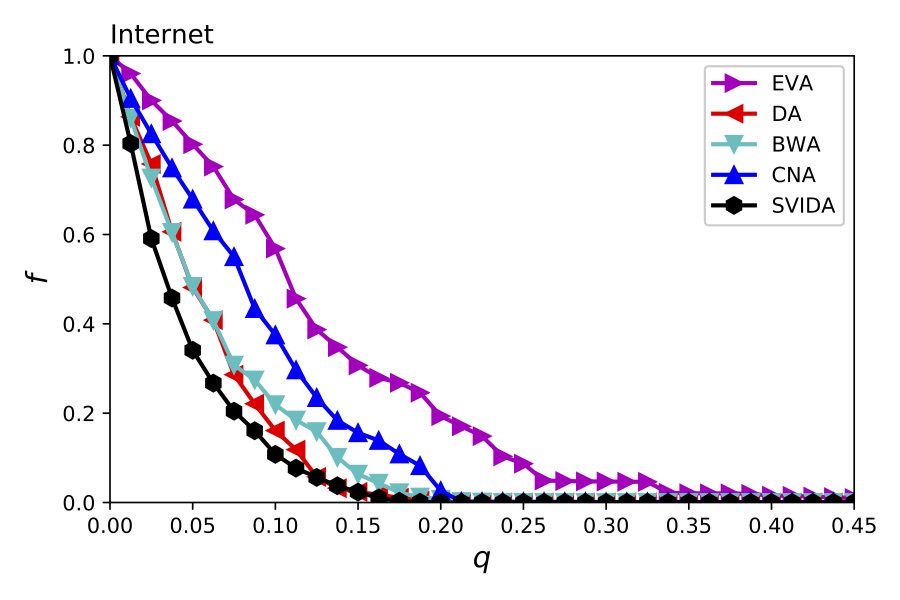}
                \caption{}
            \end{subfigure}
            \begin{subfigure}{0.49\textwidth}
                \includegraphics[width=1\linewidth,height=6cm]{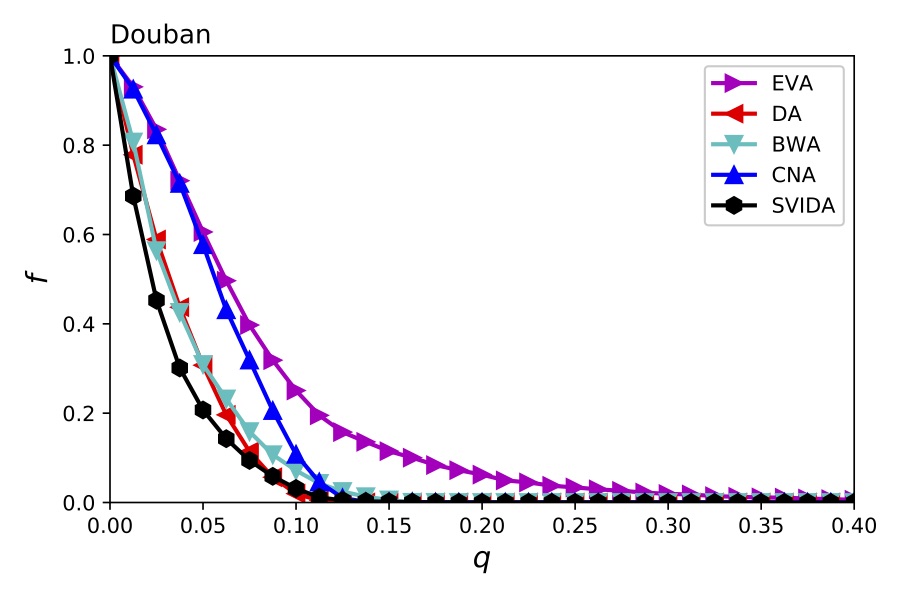}
                \caption{}
            \end{subfigure}
            
           \caption{The fraction of lcc (largest connected cluster) $f$ vs immunized nodes fraction $q$ under various strategies for the real networks.}\label{Fig. 5}           
\end{figure}
\begin{table}[ht]
\caption{The robustness(R) of different immunization methods on four empirical networks}
\begin{center}
\begin{tabularx}{\textwidth}{XXXXX}
  \hline
 Network &AS &Email &Internet &Douban\\ \hline
EVA &0.234943 &0.194880 &0.269455 &0.442894\\
DA &0.062336 &0.142614 &0.125857 &0.226204\\
BWA &0.065080 &0.156950 &0.130640 &0.243314\\
CNA &0.108387 &0.174142 &0.184121 &0.334166\\
SVIDA &0.062398 &0.130150 &0.103148 &0.184358\\  \hline
\end{tabularx}
\end{center}
\label{Table 2}
\end{table} 
\\
\hspace*{3ex}The degree centrality finds highest order nodes which estimate impact on immediate nodes and immunizing such candidates results in lowering network density which is a prominent factor to effect the growth of network contagion. Similarly nodes with highest betweenness centrality eliminate alternative paths of transmission for the spread. Immunizing nodes with higher eigenvector centrality withheld the propagation of epidemic due to removal of influential nodes. The coreness centrality identifies nodes for inoculation target which are at most risk of infection. Consider \ref{Fig. 3}(a), (c), (e), (g) with $\lambda=0.2,\sigma=0.1$, we find that immunizing strategies based on DA and BWA are comparable as it requires roughly equal fractions of immunizing nodes to completely withheld the spread in networks while strategy based on SVIDA shows up to be more effective for the similar cases. The proposed SVIDA is a local information based impelling criteria for immunization as it inclines to find the nodes among cut-vertices/bridges between components which is analogous to BWA which requires complete information on graph to calculate. Comparable to local information based criteria like DA, proposed SVIDA locates hubs nodes who are  connected to maximum numbers of subgroups in the graph. When such hubs and bridges are removed from the graph, it leads to losing number of alternative paths among different subgroups and failing communication between remaining nodes after immunization. For infect rate $\lambda= 0.8$, \ref{Fig. 3}(b), (d), (f), (h) explain the similar results and in this case greater fraction of $q$ is required to cease the epidemic spread. This is due to more strong infection that it requires more immunizing population to block the effective spreading. We further examine network vulnerability upon immunization by plotting fraction of largest connected cluster nodes $f$ versus $q$, the fraction of the immunized nodes. Using different strategies on model networks \ref{Fig. 4}, SVIDA exhibits high precedence compared with other strategies for reducing the size of lcc during the process of immunization. The DA performs close to SVIDA in case of SF network (\ref{Fig. 4}) and finally both SVIDA and DA take approx $15\%$ of the nodes,  essentially to be removed for the giant component to lose. \ref{Fig. 5} shows results of plotting $f$ vs $q$ over real-world networks. Against the next best strategies applied on different networks, the reduction in size of lcc by SVIDA shows prevalence. For AS network \ref{Fig. 5}(a) performance of DA, BWA and SVIDA is comparative and practically all three strategies establish total collapse at $q\approx0.15$. However, with various amount of immunization fraction ($q$), the resulted $f$ by SVIDA is smaller than BWA by $14\%$ and $6\%$ smaller in case of DA. For email network \ref{Fig. 5}(b) SVIDA takes $q\approx0.15$, however DA, BWA, CNA and EVA take roughly same values of $q\approx0.18-0.20$ for the network to collapse. In this case, for various amount of $q$, SVIDA shows up improvements of $9\%$, $5\%$, $23\%$ and $16\%$ over BWA, DA, EVA and CNA respectively.  For internet network ~\ref{Fig. 5}(c), BWA shows up next best strategy to DA and SVIDA, as both use approx $q\approx 0.16$ to lose giant component whereas BWA takes $q\approx0.22$ for the same. However SVIDA manifests improvements of $16\%$ and $17\%$ over BWA and DA. For douban network ~\ref{Fig. 5}(d) SVIDA and DA both require $q\approx 0.12$ corresponding to the loss of giant component, however SVIDA strategy generates least size of lcc over all proportions of immunized population and shows up and improvement of $18\%$ and $14\%$ more than BWA and DA.
To further quantify the performance of benchmark strategies, we enumerate robustness R of the networks as defined in Eq.~\eqref{eq:6}. The robustness is also defined as the area under the curve of $f-q$ as shown in \ref{Fig. 4} and \ref{Fig. 5}). The robustness R for different networks under different immunizing strategies is listed in \ref{Table 2}. Overall, SVIDA performs relatively best among all the strategies, DA is next to SVIDA, which shows hub nodes are very important for network connectivity. Further, BWA which finds nodes with high controllability on network flow performs better next to DA.
\section{Conclusion}\label{8} In this work we propose a novel group based measure for centrality to solve node immunization problem in graphs. Motivated from the cooperative game theoretic approach over network centrality and efficient computation of Shapley value, we approximate closed form for the game defined for information delimitation. We address the issue of defining criteria for finding bridging points in the network, such that removing these points withheld communication between the other nodes belonging to different subgroups of the graph. SVID algorithm formulates legitimate ranking between hubs and articulation points based on neighboring information of nodes connected by an edge, as a suggested future direction from the work from Narayanam and Narahari~\cite{Narayanam_2011}. We verified the performance of our proposed method based on the criteria of network vulnerability and robustness. Targeted immunization of delimiters defined by SVID caused significant abridgment of alternative paths among subgroups in the graph, leading to least network robustness throughout. However, for the same fraction of top ranked nodes removed, SVID separates singletons in the process of network disruption, the size of lcc remains lesser than benchmark methods. Finally, It is examined that the conclusion of epidemic spreading on immunized network using SVID algorithm is effective when investigated with correlated strategies.     

\bibliographystyle{plain}
\bibliography{ms}

\end{document}